\documentclass[prb,aps,showpacs,twocolumn,floatfix]{revtex4-1}
\usepackage{graphicx}
\usepackage{dcolumn}
\usepackage{bm}

\begin{document}

\title{Anhamonic finite temperature effects on the Raman and Infrared spectra to determine the crystal structure phase III of solid molecular hydrogen}
\author{Ranber Singh}
\affiliation{Institute for Physical Chemistry, Johannes Gutenberg University Mainz, Staudinger Weg 9, D-55128 Mainz, Germany}
\author{Sam Azadi}
\affiliation{Thomas Young Center and Department of Physics, Imperial College London, London SW7 2AZ, United Kingom}
\author{Thomas D. K\"{u}hne}
\email{kuehne@uni-mainz.de}
\affiliation{Institute for Physical Chemistry and Center for Computational Sciences, \\
Johannes Gutenberg University Mainz, Staudinger Weg 7, D-55128 Mainz, Germany}

\date{\today}
\begin{abstract}
We present theoretical calculations of the Raman and IR spectra, as well as electronic properties at zero and finite temperature to elucidate the crystal structure of phase III of solid molecular hydrogen. 
We find that anharmonic finite temperature are particularly important and qualitatively influences the main conclusions. While P6$_3$/m is the most likely candidate for phase III at the nuclear ground state, at finite temperature the C2/c structure appears to be more suitable. 
\end{abstract}

\date{\today}
\maketitle

\section{introduction}
The high-pressure phase of solid molecular hydrogen is very complex. Nevertheless, due to its relevance to astrophysics \cite{Alavi01091995}, the possible existence of high-$T_c$ superconductivity \cite{Ashcroft68} and a metallic quantum fluid ground state \cite{Bonev04}, the importance to grasp metallic hydrogen can hardly be overstated. However, since Hydrogen atoms are very light, their x-ray scattering is extremely weak, leading to extremely low-resolution diffraction patterns \cite{Mao94,Loubeyre96}. This makes it particularly challenging to determine the crystalline structure of solid hydrogen at high pressure. Experiments based on vibrational techniques, such as Raman and infrared (IR) spectroscopies, have identified four different solid phases (I, II, III and IV) of molecular hydrogen at megabar pressures (100$-$300 GPa) and different temperature ranges \cite{Mao94,Eremets11,Howie12}. The low pressure phase ($\leq$ 110 GPa) is denoted as phase I and consists of freely rotating hydrogen molecules on an hexagonal close-packed lattice. The low temperature structures at pressures above 110 GPa, known as phase II and III, are still unknown \cite{PhysRevLett.61.857, PhysRevLett.63.2080}, just like the recently discovered phase IV that is stable at temperatures well above the nuclear ground state \cite{Howie12, PhysRevLett.110.217402}. This is due to the fact that the combination of scattering and vibrational spectroscopy measurements only reveal the state of bonding of the hydrogen atoms, but are not sufficient to uniquely determine the corresponding crystal structures. 
\par 
Even though it is generally still impossible to ensure that a particular structure corresponds to the global minimum in enthalpy, recently great strides have been made to predict crystal structures from first-principles, 
to the extend that allowed Pickard and Needs to systematically investigate the zero-temperature phase diagram of solid molecular hydrogen by means of theory. 
In particular, they suggested the C2/c phase to be energetically most favorable for the pressure range of phase III \cite{Pickard07}. At variance, based on more accurate hybrid density functional theory (DFT) calculations that includes some fraction of exact Hartree-Fock exchange, Azadi and K\"{u}hne predicted the P6$_3$/m to be the most likely candidate for phase III of solid hydrogen \cite{Azadi12}. Even more accurate Quantum Monte Carlo calculations, including the harmonic part of the zero-point energy (ZPE), also indicates that the P6$_3$/m phase obeys the lowest enthalpy up to 220 GPa, whereas for higher pressures the C2/c structure is enthalpically most favorable \cite{Azadi13}. 
However, the differences in enthalpy within these structures are very small and nuclear quantum effects (NQE), such as the ZPE, relatively large, which puts rather stringent accuracy requirements on the calculation of both contributions. While the former is typically calculated using DFT, the latter is often estimated using the quasi-harmonic approximation. Yet both of them have recently been shown to be not accurate enough for the reliable prediction of high-pressure crystal structures. On the one hand Azadi et al. have shown that an electronic structure theory beyond DFT is essential \cite{AzadiFoulkes13, Azadi13}, whereas on the other hand Ceperley and coworkers demonstrated the importance of anharmonic ZPE and the significance of finite temperature effects \cite{Morales13b}. Unfortunately, combining path-integral molecular dynamics, which exactly accounts for finite temperature and NQE, with an accurate electronic structure technique beyond DFT is computational still not feasible. 
\par
Therefore, in this paper we follow a different approach and consider anharmonic finite temperature effects by means of DFT-based Born-Oppenheimer molecular dynamics (BOMD) simulations at room temperature \cite{RevModPhys.64.1045, CP2G}. However, we do not attempt here to determine the most likely candidate structure in the pressure range of phase III by averaging the enthalpy from accurate electronic structure calculations.
Instead, we calculate here the ensemble averages of the Raman and IR spectra, which are much more sensitive to structural changes, and compare them directly to experimental measurements at room temperature to assign the structure of phase III of solid molecular hydrogen \cite{Akahama10, Zha12}.

\section{computational details}
We considered here the C2, C2/c, Cmca-12, Pbcn and P6$_3$/m phases, which have all been found to be enthalpically competitive candidate structures for phase III of solid molecular hydrogen \cite{Pickard07}. The supercell of the latter structure consisted of 64 hydrogen molecules, whereas all the others supercells contained 48 hydrogen molecules, respectively. All of our DFT calculations were performed using the projector augmented-wave method, as implemented in Quantum Espresso \cite{PhysRevB.50.17953, Giannozzi09}. We employed the Perdew-Burke-Ernzerhof (PBE) generalized gradient approximation to the exchange and correlation energy \cite{Perdew96}. The BOMD simulations were conducted in the isobaric-isothermal NPT ensemble, i.e. at constant pressure (250~GPa) and temperature (200~K and 300~K). Each of the simulations were initially equilibrated for 5~ps before accumulating statistics for another 15~ps using an integration timestep of 8 a.u. (1 a.u.= 0.0484~fs). 
It is important to note that in this way NQE are neglected. Nevertheless, since NQE are at finite temperature much smaller than at the nuclear ground-state and furthermore primarily influence the relative stability of the various crystal phases rather than the Raman and IR spectra, the effect on the eventual structure determination can be assumed to be small.  
\par
For the purpose to calculate the nuclear forces for the BOMD simulation, we have employed a plane wave basis set with an energy cut-off of 680~eV and a 2$\times$2$\times$2 Monkhorst-Pack k-point mesh to sample the first Brillouin zone \cite{PhysRevB.13.5188}. Beside calculating the Raman and IR spectra, the electronic density of states (EDOS) and the electronic band gaps at the nuclear ground state at 0~K, we have also calculated them at finite temperature as ensemble averages of 200 equidistantly selected configurations from our BOMD simulations. To that extend we used an energy cutoff of 816 eV and a dense 18$^3$ k-point mesh to compute the EDOS and the band gaps, whereas a rather hard norm-conserving (NC) pseudopotential \cite{PhysRevB.58.3641} with a cutoff of 1360~eV and the Perdew-Zunger (PZ81) exchange and correlation functional \cite{PhysRevB.23.5048} had been employed to calculate the Raman and IR spectra. Optimized structures were obtained through minimizing the enthalpy when simultaneously varying the atomic positions and the simulation cell by a dynamical simulated annealing \cite{PhysRevLett.55.2471, PhysRevLett.98.066401}. 
The spectra were all calculated at the $\Gamma$-point only using density-functional perturbation theory and its third-order extension based on the `$2n+1$' rule to compute the anharmonic cubic coupling constants \cite{Debernardi1994813, RevModPhys.73.515}.

\section{results and discussion}
The main constraints that are used to identify the various phases of solid molecular hydrogen comes from experimental vibrational spectroscopy measurements. The lattice vibration modes due to low-frequency intermolecular interactions are denoted as librons, while the high-frequency intramolecular stretching modes are known as vibrons. The pressure induced transition from phase II to phase III can be identified in the Raman spectrum by a large discontinuous drop of the Raman vibron frequency \cite{Loubeyre02,Akahama10a} followed by a continuous decrease upon further increase in pressure \cite{Loubeyre02,Akahama10,Zha12}. This indicates not only strong intermolecular interactions, but also the presence of orientational ordering in phase III. Furthermore, phase III obeys multiple low-frequency Raman-active modes \cite{Loubeyre02}, whereas the frequency of the highest Raman-active libron increases with pressure \cite{Goncharov01}. 
However, it is also possible to detect the pressure induced phase II to III transition within experimental IR spectra by a discontinuous drop in frequency of the IR-active vibrons \cite{Hanfland93}. Increasing the pressure even further leads to a large increase in the intensity of the IR-active vibrons \cite{Hanfland93,Mazin97} and to a decrease in the frequency of the IR-active vibrons \cite{Goncharov01,Zha12,Hanfland93}. The distinction between {\it para} and {\it ortho} hydrogen molecules is herein of minor importance, due to the rather small rotational freedom of the individual molecules \cite{Mazin97}.
\par
All of the structures considered here consists of hydrogen molecules arranged in layers, though each time with a different stacking. In any case, the centers of the molecules lie on a distorted hexagonal lattice. The individual layers of the P6$_3$/m and Cmca-12 phases are stacked in a ABA fashion, whereas the C2/c structure obeys a ABCDA stacking. At variance, the C2 and Pbcn phases are mixed layered structures made of alternating layers of strongly and weakly bonded hydrogen molecules. However, initial Raman and IR spectra calculations immediately suggest that the latter mixed layered phases are inconsistent with phase III of solid molecular hydrogen. Henceforth, we will confine ourselves to the C2/c, P6$_3$/m and Cmca-12 phases only and will discuss the C2 and Pbcn structures in a forthcoming paper on phase IV of solid hydrogen \cite{SinghPhaseIV}. 
Except for the P6$_3$/m phase, where three-quarters of the hydrogen molecules are planar, while one quarter of the molecules are oriented perpendicular to the plane, in all other structures the hydrogen molecules are arranged in-plane.

\subsection{Raman and IR spectra at the nuclear ground state}
\begin{figure}[htb]
\begin{centering}
\includegraphics[width=\linewidth]{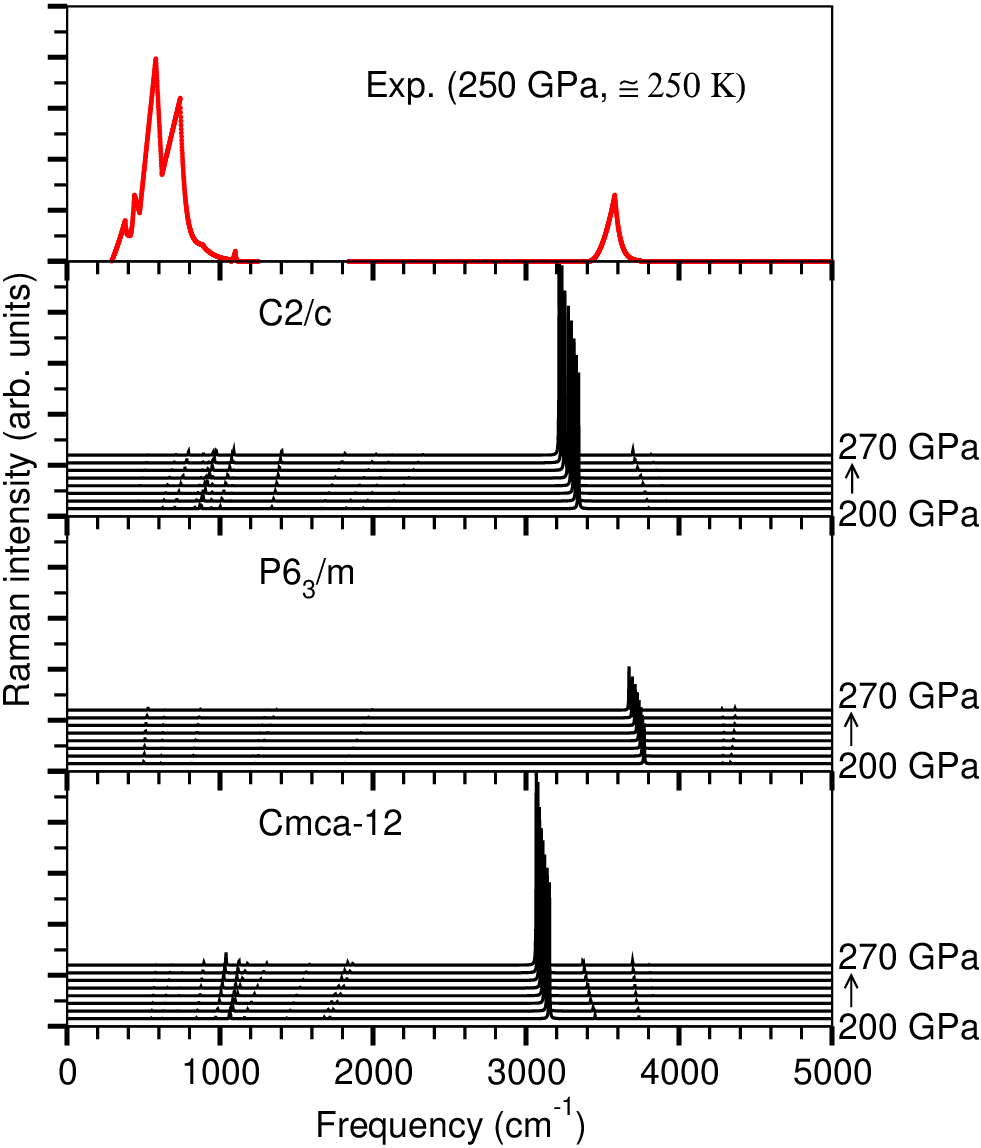}
\caption{\label{fig:Raman-0K} The calculated Raman spectra of the C2/c, P6$_3$/m and Cmca-12 crystal structures of solid molecular hydrogen as a function of pressure at 0~K. The experimental data in the top panel is adapted from Ref.~\onlinecite{Howie12a}.} 
\end{centering}
\end{figure}

The calculated Raman spectra as a function of pressure of the C2/c, P6$_3$/m and Cmca-12 phases at their respective nuclear ground states are shown in Fig.~\ref{fig:Raman-0K}. 
In comparison to the P6$_3$/m phase, the intensity of the Raman-active modes of the C2/c and Cmca-12 structures is about three times stronger. In all of the considered structures, the frequencies of the Raman-active libron modes are increasing, while the frequencies of the strong Raman-active vibron modes are decreasing with pressure, which is in qualitative agreement with the experimental data \cite{Loubeyre02,Akahama10,Zha12,Goncharov01}. 
However, the P6$_3$/m phase, which exhibits three Raman-active vibron modes, is the sole configuration that is in agreement with a previous group theoretical analysis \cite{Cui95} and is overall in best agreement with the experimental Raman spectra \cite{Howie12a} as can be seen in Fig.~\ref{fig:Raman-0K}. 
\par
\begin{figure}[htb]
\begin{centering}
\includegraphics[width=\linewidth]{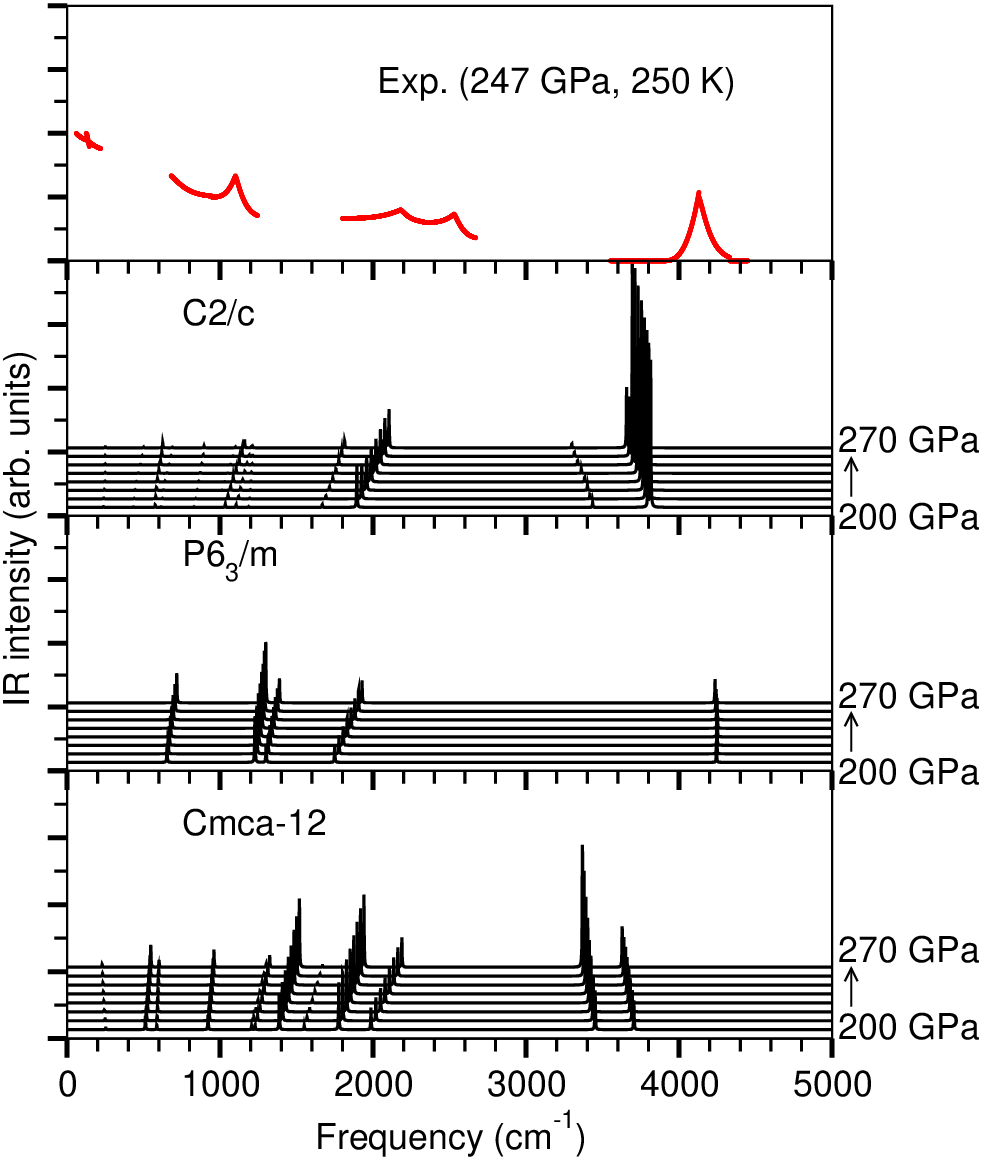}
\caption{\label{fig:IR-0K} The calculated IR spectra of the C2/c, P6$_3$/m and Cmca-12 crystal structures of solid molecular hydrogen as a function of pressure at 0~K. The experimental data in the top panel is adapted from Ref.~\onlinecite{Zha12}.}
\end{centering}
\end{figure}

The corresponding IR spectra of the very same crystal structures as a function of pressure at 0~K are given in Fig.~\ref{fig:IR-0K}. In comparison with the P6$_3$/m phase, the intensity of the dominant IR vibron mode of the C2/c structure is about five times stronger, and almost twice as big as the Cmca-12 one. Upon increasing the pressure in all of the examined structures, the frequencies of the IR-active libron modes are increasing, while the frequencies of IR-active vibrons are decreasing, which is qualitatively in agreement with the experimental data \cite{Zha12,Goncharov01}. However, experimentally only one IR-active vibron mode is observed, whereas the C2/c and Cmca-12 phases have each two IR-active vibron modes. Thus, the P6$_3$/m structure, which possess only one IR-active vibron mode, is the only composition that is in agreement with an earlier group theoretical analysis \cite{Cui95}. Even though, looking at Fig.~\ref{fig:IR-0K}, it is apparent that the P6$_3$/m phase is in best agreement with the experimental IR measurements, the marginally decreasing frequency of the IR vibron mode with respect to pressure is at variance to the experimental findings, whose IR vibron frequency is noticeably decreasing with increasing pressure \cite{Zha12}. 
As a consequence, beside the P6$_3$/m phase, the C2/c structure, which not only has a rather large IR vibron intensity \cite{Hanfland93,Goncharov01}, but furthermore an with pressure decreasing IR frequency \cite{Zha12,Goncharov01}, are the two most likely candidates for phase III of solid molecular hydrogen. 
\par
\begin{figure}[htb]
\begin{centering}
\includegraphics[width=\linewidth]{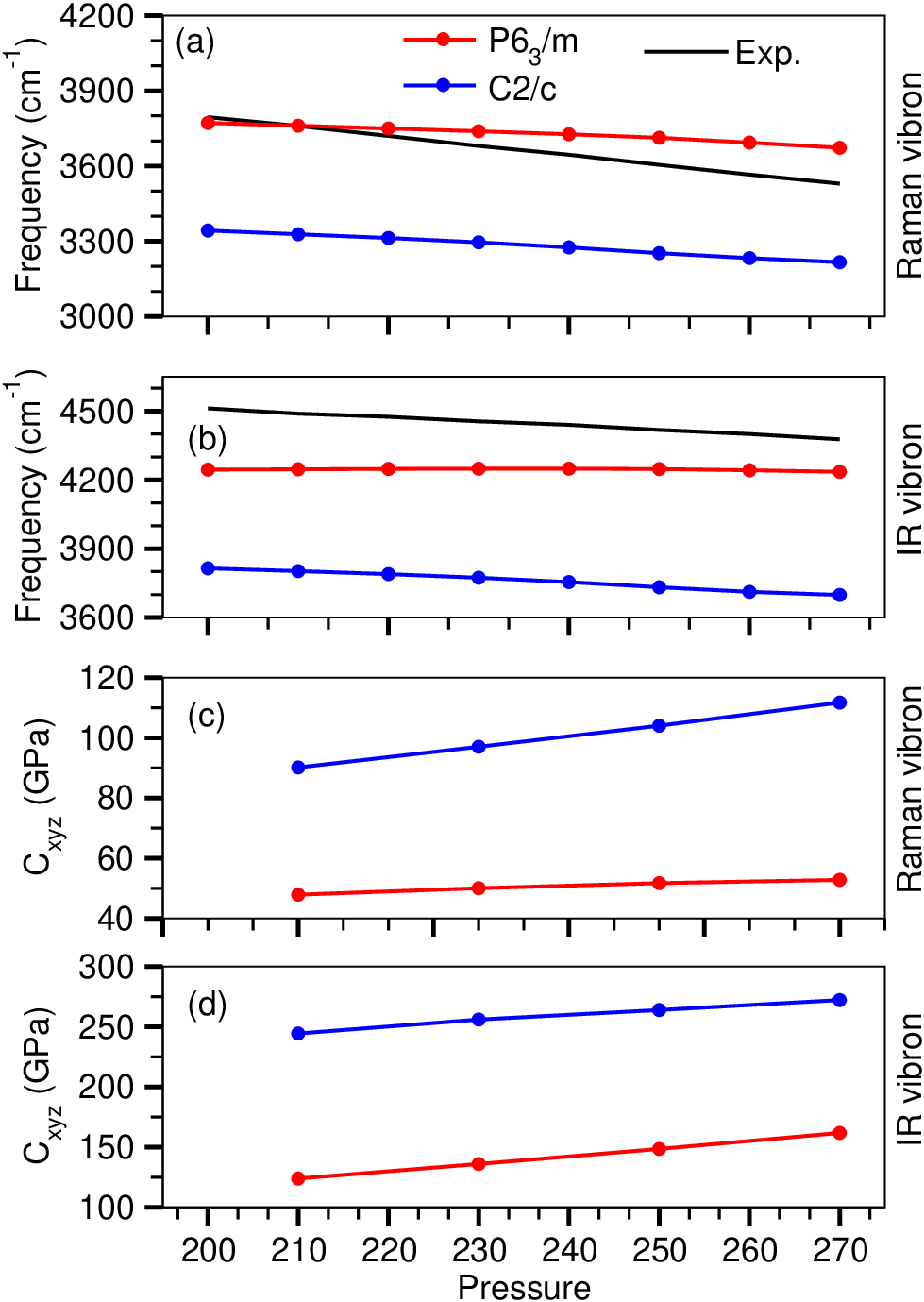}
\caption{\label{fig:Vibron} The strongest Raman- and IR-active vibron frequencies as well as the third-order anharmonic force constants of the C2/c and P6$_3$/m phases as a function of pressure. The experimental data is adapted from Ref.~\onlinecite{Zha12}. The anharmonic cubic coupling constants are denoted as $\text{C}_{\text{xyz}}$.}
\end{centering}
\end{figure}

 
In Fig.~\ref{fig:Vibron} the calculated frequencies as a function of pressure of the strongest Raman- and IR-active vibron modes of the remaining P6$_3$/m and C2/c phases are compared with the experimental ones. 
As can be seen in Fig.~\ref{fig:Vibron}a, both candidate structures exhibit the correct slope of the strongest mode with respect to pressure, whereas for the P6$_3$/m phase only, in addition the frequency itself is in excellent agreement with experiment \cite{Zha12}. 
On the contrary, the assignment based on the IR-active vibron modes of Fig.~\ref{fig:Vibron}b is less obvious. While the P6$_3$/m phase is to favor regarding the quantitative agreement with experiment of the strongest IR vibron frequency, its slope as a function of pressure is better reproduced by the C2/c structure. 
\par
Furthermore, we have calculated the third-order anharmonic force constants associated with the strongest Raman- and IR-active vibrons, respectively. 
From Fig.~\ref{fig:Vibron}c-d it is evident that the cubic force constants of both phases are increasing with pressure, which is most likely due to the increasing anharmonicity of the lattice vibrations as a function of pressure. 
This immediately suggests that the linewidth of the Raman- and IR-active vibron peaks is increasing with pressure as well, in agreement with the experimental data of phase III of solid molecular hydrogen \cite{Howie12,Zha12}. 

\subsection{Raman and IR spectra at finite temperature}
\begin{figure}[htb]
\begin{centering}
\includegraphics[width=\linewidth]{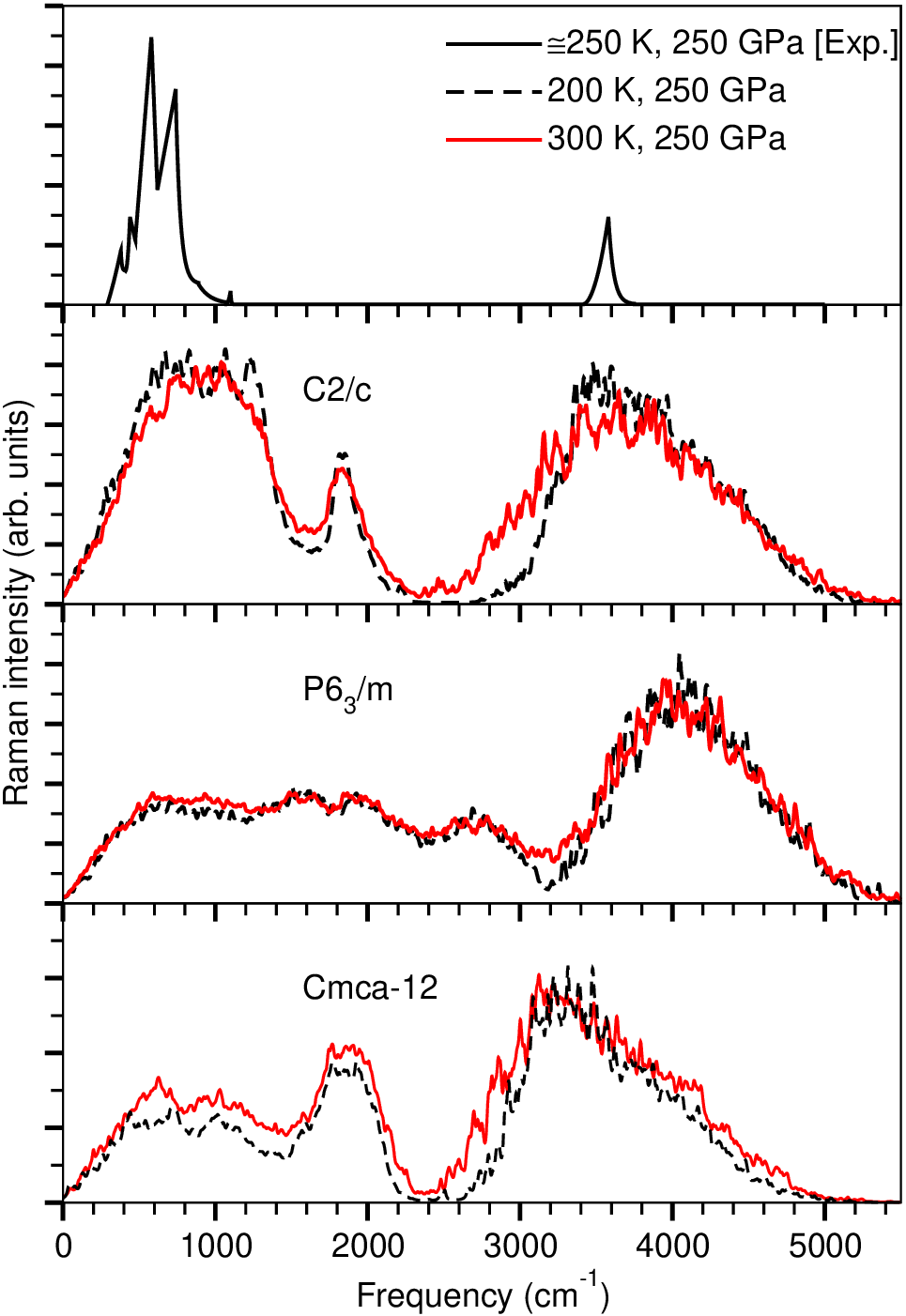}
\caption{\label{fig:Raman-300K} The calculated Raman spectra of the C2/c, P6$_3$/m and Cmca-12 crystal structures of solid molecular hydrogen at temperatures of 200~K and 300~K and a pressure of 250~GPa. The experimental data in the top panel is adapted from Ref.~\onlinecite{Howie12a}.}
\end{centering}
\end{figure}

\begin{figure}[htb]
\begin{centering}
\includegraphics[width=\linewidth]{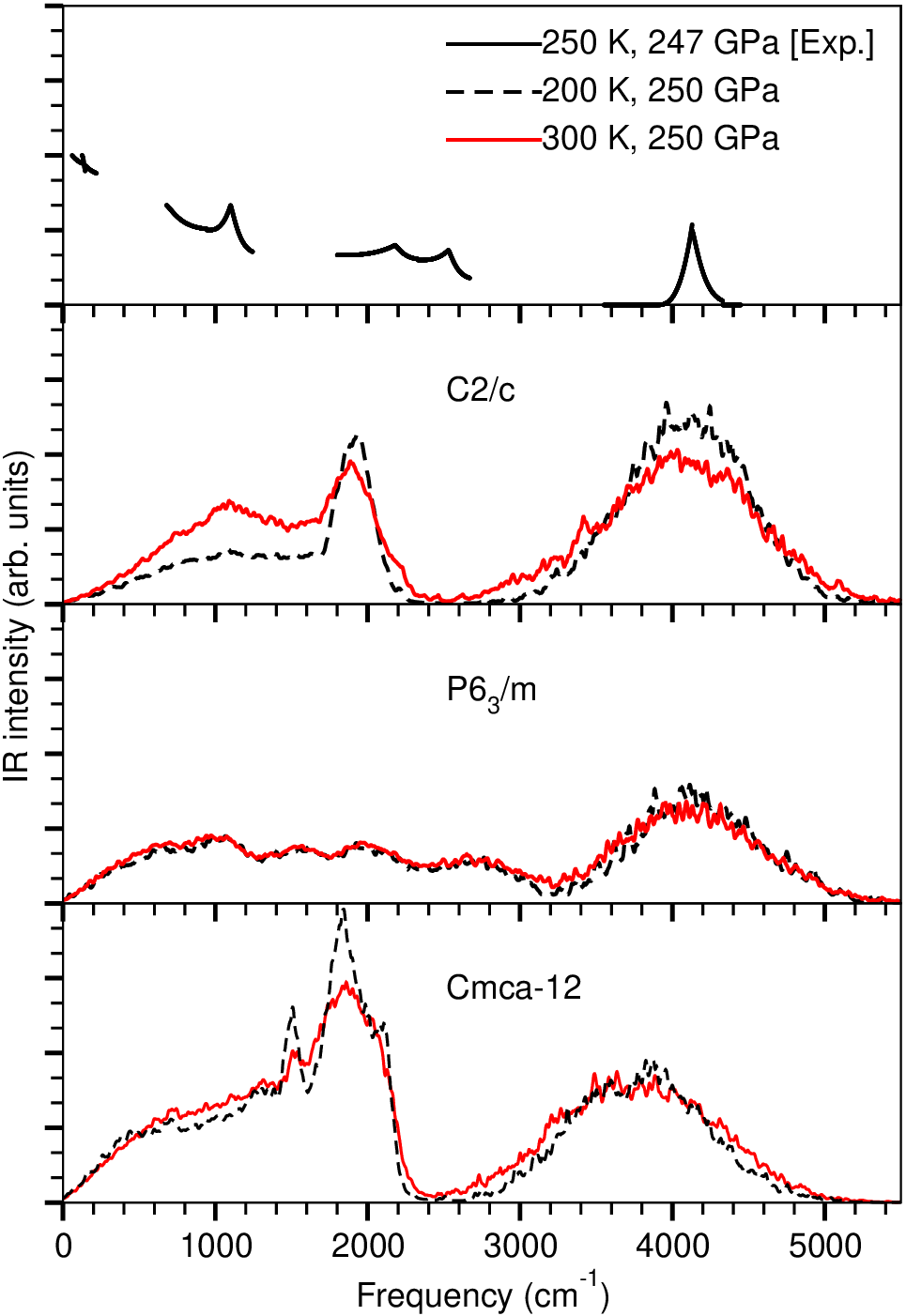}
\caption{\label{fig:IR-300K} The calculated IR spectra of C2/c, P6$_3$/m and Cmca-12 crystal structures of solid molecular hydrogen at temperatures of 200~K and 300~K and a pressure of 250~GPa. The experimental data in the top panel is adapted from Ref.~\onlinecite{Zha12}.}    
\end{centering}
\end{figure}

The calculated Raman and IR spectra of the C2/c, P6$_3$/m and Cmca-12 crystal structures that comply with nuclear temperatures of 200~K and 300~K and a pressure of 250~GPa are given in Figs.~\ref{fig:Raman-300K} and \ref{fig:IR-300K}, respectively. The most apparent distinction from the corresponding spectra at 0~K is the expected broadening of all the peaks due to anharmonic finite temperature. Nevertheless, comparing the actual Raman spectra at finite temperature with the ones of Fig.~\ref{fig:Raman-0K} at 0~K unveils that the Raman-active vibron modes are all experiencing an anharmonic finite temperature induced redshift. The fact that the redshift amount to 200-300~cm$^{-1}$ is in effect rather consequential and entails that at finite temperature the C2/c structure is now in better agreement with the experimental data than the P6$_3$/m phase, which was the closest match at 0~K \cite{Howie12a}. In addition, the low-frequency librons of the C2/c structure are in qualitative agreement with the experimental ones, while the agreement of the P6$_3$/m phase is rather poor. 
However, the linewidth of the Raman-active vibron of all probed crystal structures is very large compared with the experimental value of about 100 cm$^{-1}$, which is most likely due to an overestimation of the anharmonic proton motion by the present semi-classical BOMD simulations. \\
\par 
On the contrary, the influence of anharmonic finite temperature effects on the IR spectra shown in Fig.~\ref{fig:IR-300K} is somewhat more subtle. Whereas the IR vibron mode of the C2/c and Cmca-12 structures is again redshifted by about 300~cm$^{-1}$, the P6$_3$/m phase experiences an anharmonic finite temperature induced blueshift of about 150~cm$^{-1}$. Even though the IR-active vibrons of both, the C2/c and the P6$_3$/m configurations are now consistent with the experimental measured ones, the agreement of the libron modes is generally less satisfying and just slightly in favor of the C2/c structure \cite{Zha12}.
Similar to the Raman spectra, the linewidth of the IR-active vibron is very large compared to the experimental value. 
\par
For all of the examined crystal structures, the Raman as well as the IR spectra were rather similar for 200~K and 300~K, respectively, except for a minor increase in the linewidth of the vibron modes. However, at variance to others, no temperature induced phase transition has been observed for any of the considered phases \cite{Liu13,Magdau13}.

\subsection{Electronic properties}
From Table~\ref{tab:bandgap} it can be deduced that anharmonic finite temperature effects reduces the electronic band gaps of the C2/c and P6$_3$/m phases by 0.63~eV and 2.13~eV, respectively. On the contrary, the energy gap of the Cmca-12 structure is increased by 0.66~eV, which is another manifestation that the inclusion of anharmonic finite temperature effects is essential. In all cases the computed band gaps turn out to be indirect energy gaps only. At 0~K the well known rule "the lower the energy, the wide the gap" \cite{PhysRevLett.67.1138} is fulfilled and the P6$_3$/m structure indeed exhibits the largest gap. At finite temperature, however, the Cmca-12 phase breaks ranks. Anyhow, the roughly 1~eV smaller band gap of the C2/c composition in comparison to the P6$_3$/m phase, can be thought of as a further indication in support of the C2/c structure as the most likely candidate for phase III of solid molecular hydrogen at finite temperature.
\begin{table}[htb]
\caption{The electronic gaps of the C2/c, P6$_3$/m and Cmca-12 crystal structures of solid molecular hydrogen at temperatures of 0~K and 300~K and a pressure of 250~GPa..} 
\centering
\begin{ruledtabular}
\begin{tabular}{c|c|c}
 Crystal         &  \multicolumn{2}{c}{Electronic Band Gaps} \\
Structure & Nuclear Ground State  & Finite Temperature   \\\hline
C2/c      & 0.65~eV & 0.02~eV    \\
P6$_3$/m  & 1.08~eV & -1.05~eV  \\
Cmca-12   & -0.61~eV & 0.05~eV
\end{tabular}
\end{ruledtabular}
\label{tab:bandgap}
\end{table}
\par
\begin{figure}[htb]
\begin{centering}
\includegraphics[width=\linewidth]{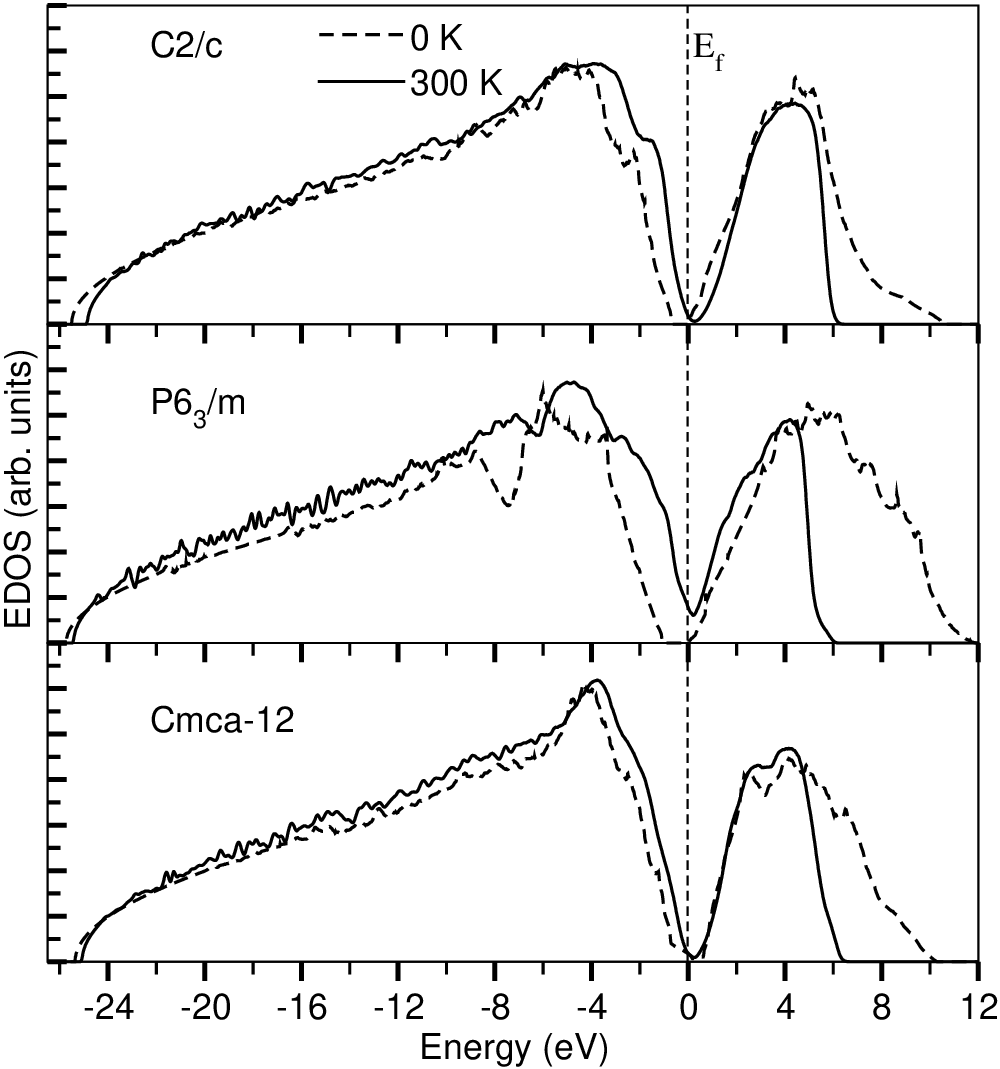}
\caption{\label{fig:EDOS} The EDOS of the C2/c, P6$_3$/m and Cmca-12 crystal structures of solid molecular hydrogen. The solid (dotted) lines represent the EDOS at 300~K (0~K) temperatures and a pressure of 250~GPa. The EDOS has been computed setting the Fermi level energy (E$_f$) equal to zero.}
\end{centering}
\end{figure}

Furthermore, in Fig.~\ref{fig:EDOS} the associated EDOS are given. As can be seen, the P6$_3$/m phase is the sole configuration that has a non-zero EDOS around the Fermi level at 300~K. All the other structures had zero EDOS around the Fermi level at 0~K and 300~K, respectively. 
However, due to the fact that even the exact single-particle Kohn-Sham gap differs from the true fundamental gap by the infamous derivative discontinuity, local and semi-local DFT often severely underestimates the electronic band gap \cite{PhysRevLett.51.1884}. Taking into account that typical gradient corrected exchange and correlation functionals underestimate the true band gap by at least 1-2~eV \cite{Stadele00, Azadi12, Azadi13}, all of the investigated candidate structures for phase III of solid molecular hydrogen are expected to be insulating at the nuclear ground state as well as at room temperature, which is in agreement with the experimental results \cite{Zha12}.

\section{Conclusion}
We have presented theoretical calculations of the Raman and IR spectra, as well as electronic properties at zero and finite temperature on various low-enthalpy crystal structures, which have been proposed as potential candidates for the phase III of solid molecular hydrogen. 
Among the various candidate structures we have investigated, we found that at 0~K the P6$_3$/m phase, which has been recently predicted by K\"uhne and coworkers \cite{Azadi12, Azadi13}, is indeed in best agreement with the experimental data of phase III of solid molecular hydrogen. However, at finite temperature the C2/c structure appeared to be the most likely candidate for phase III, which clearly demonstrates the importance of anharmonic finite temperature effects that have the potential to qualitatively alter the main conclusions.
The fact that all of the examined crystal structures were insulating immediately excludes the theoretically conjectured existence of phonon-driven superconductivity in phase III of solid molecular hydrogen \cite{HydrogenHTC, PhysRevLett.78.118}. 

\begin{acknowledgements}
We would like to acknowledge financial support from the Graduate School of Excellence MAINZ and IDEE project of the Carl Zeiss Foundation.
\end{acknowledgements}


\begin{thebibliography}{44}%
\makeatletter
\providecommand \@ifxundefined [1]{%
 \@ifx{#1\undefined}
}%
\providecommand \@ifnum [1]{%
 \ifnum #1\expandafter \@firstoftwo
 \else \expandafter \@secondoftwo
 \fi
}%
\providecommand \@ifx [1]{%
 \ifx #1\expandafter \@firstoftwo
 \else \expandafter \@secondoftwo
 \fi
}%
\providecommand \natexlab [1]{#1}%
\providecommand \enquote  [1]{``#1''}%
\providecommand \bibnamefont  [1]{#1}%
\providecommand \bibfnamefont [1]{#1}%
\providecommand \citenamefont [1]{#1}%
\providecommand \href@noop [0]{\@secondoftwo}%
\providecommand \href [0]{\begingroup \@sanitize@url \@href}%
\providecommand \@href[1]{\@@startlink{#1}\@@href}%
\providecommand \@@href[1]{\endgroup#1\@@endlink}%
\providecommand \@sanitize@url [0]{\catcode `\\12\catcode `\$12\catcode
  `\&12\catcode `\#12\catcode `\^12\catcode `\_12\catcode `\%12\relax}%
\providecommand \@@startlink[1]{}%
\providecommand \@@endlink[0]{}%
\providecommand \url  [0]{\begingroup\@sanitize@url \@url }%
\providecommand \@url [1]{\endgroup\@href {#1}{\urlprefix }}%
\providecommand \urlprefix  [0]{URL }%
\providecommand \Eprint [0]{\href }%
\providecommand \doibase [0]{http://dx.doi.org/}%
\providecommand \selectlanguage [0]{\@gobble}%
\providecommand \bibinfo  [0]{\@secondoftwo}%
\providecommand \bibfield  [0]{\@secondoftwo}%
\providecommand \translation [1]{[#1]}%
\providecommand \BibitemOpen [0]{}%
\providecommand \bibitemStop [0]{}%
\providecommand \bibitemNoStop [0]{.\EOS\space}%
\providecommand \EOS [0]{\spacefactor3000\relax}%
\providecommand \BibitemShut  [1]{\csname bibitem#1\endcsname}%
\let\auto@bib@innerbib\@empty
\bibitem [{\citenamefont {Alavi}\ \emph {et~al.}(1995)\citenamefont {Alavi},
  \citenamefont {Parrinello},\ and\ \citenamefont {Frenkel}}]{Alavi01091995}%
  \BibitemOpen
  \bibfield  {author} {\bibinfo {author} {\bibfnamefont {A.}~\bibnamefont
  {Alavi}}, \bibinfo {author} {\bibfnamefont {M.}~\bibnamefont {Parrinello}}, \
  and\ \bibinfo {author} {\bibfnamefont {D.}~\bibnamefont {Frenkel}},\ }\href
  {\doibase 10.1126/science.7652571} {\bibfield  {journal} {\bibinfo  {journal}
  {Science}\ }\textbf {\bibinfo {volume} {269}},\ \bibinfo {pages} {1252}
  (\bibinfo {year} {1995})}\BibitemShut {NoStop}%
\bibitem [{\citenamefont {Ashcroft}(1968)}]{Ashcroft68}%
  \BibitemOpen
  \bibfield  {author} {\bibinfo {author} {\bibfnamefont {N.~W.}\ \bibnamefont
  {Ashcroft}},\ }\href {\doibase 10.1103/PhysRevLett.21.1748} {\bibfield
  {journal} {\bibinfo  {journal} {Phys. Rev. Lett.}\ }\textbf {\bibinfo
  {volume} {21}},\ \bibinfo {pages} {1748} (\bibinfo {year}
  {1968})}\BibitemShut {NoStop}%
\bibitem [{\citenamefont {Bonev}\ \emph {et~al.}(2004)\citenamefont {Bonev},
  \citenamefont {Schwegler}, \citenamefont {Ogitsu},\ and\ \citenamefont
  {Galli}}]{Bonev04}%
  \BibitemOpen
  \bibfield  {author} {\bibinfo {author} {\bibfnamefont {S.~A.}\ \bibnamefont
  {Bonev}}, \bibinfo {author} {\bibfnamefont {E.}~\bibnamefont {Schwegler}},
  \bibinfo {author} {\bibfnamefont {T.}~\bibnamefont {Ogitsu}}, \ and\ \bibinfo
  {author} {\bibfnamefont {G.}~\bibnamefont {Galli}},\ }\href
  {http://dx.doi.org/10.1038/nature02968} {\bibfield  {journal} {\bibinfo
  {journal} {Nature}\ }\textbf {\bibinfo {volume} {431}},\ \bibinfo {pages}
  {669} (\bibinfo {year} {2004})}\BibitemShut {NoStop}%
\bibitem [{\citenamefont {Mao}\ and\ \citenamefont {Hemley}(1994)}]{Mao94}%
  \BibitemOpen
  \bibfield  {author} {\bibinfo {author} {\bibfnamefont {H.-K.}\ \bibnamefont
  {Mao}}\ and\ \bibinfo {author} {\bibfnamefont {R.~J.}\ \bibnamefont
  {Hemley}},\ }\href {\doibase 10.1103/RevModPhys.66.671} {\bibfield  {journal}
  {\bibinfo  {journal} {Rev. Mod. Phys.}\ }\textbf {\bibinfo {volume} {66}},\
  \bibinfo {pages} {671} (\bibinfo {year} {1994})}\BibitemShut {NoStop}%
\bibitem [{\citenamefont {Loubeyre}\ \emph {et~al.}(1996)\citenamefont
  {Loubeyre}, \citenamefont {LeToullec}, \citenamefont {Hausermann},
  \citenamefont {Hanfland}, \citenamefont {Hemley}, \citenamefont {Mao},\ and\
  \citenamefont {Finger}}]{Loubeyre96}%
  \BibitemOpen
  \bibfield  {author} {\bibinfo {author} {\bibfnamefont {P.}~\bibnamefont
  {Loubeyre}}, \bibinfo {author} {\bibfnamefont {R.}~\bibnamefont {LeToullec}},
  \bibinfo {author} {\bibfnamefont {D.}~\bibnamefont {Hausermann}}, \bibinfo
  {author} {\bibfnamefont {M.}~\bibnamefont {Hanfland}}, \bibinfo {author}
  {\bibfnamefont {R.~J.}\ \bibnamefont {Hemley}}, \bibinfo {author}
  {\bibfnamefont {H.-K.}\ \bibnamefont {Mao}}, \ and\ \bibinfo {author}
  {\bibfnamefont {L.~W.}\ \bibnamefont {Finger}},\ }\href@noop {} {\bibfield
  {journal} {\bibinfo  {journal} {Nature}\ }\textbf {\bibinfo {volume} {383}},\
  \bibinfo {pages} {702} (\bibinfo {year} {1996})}\BibitemShut {NoStop}%
\bibitem [{\citenamefont {Eremets}\ and\ \citenamefont
  {Troyan}(2011)}]{Eremets11}%
  \BibitemOpen
  \bibfield  {author} {\bibinfo {author} {\bibfnamefont {M.~I.}\ \bibnamefont
  {Eremets}}\ and\ \bibinfo {author} {\bibfnamefont {I.~A.}\ \bibnamefont
  {Troyan}},\ }\href@noop {} {\bibfield  {journal} {\bibinfo  {journal} {Nature
  Mater.}\ }\textbf {\bibinfo {volume} {10}},\ \bibinfo {pages} {927} (\bibinfo
  {year} {2011})}\BibitemShut {NoStop}%
\bibitem [{\citenamefont {Howie}\ \emph
  {et~al.}(2012{\natexlab{a}})\citenamefont {Howie}, \citenamefont {Guillaume},
  \citenamefont {Scheler}, \citenamefont {Goncharov},\ and\ \citenamefont
  {Gregoryanz}}]{Howie12}%
  \BibitemOpen
  \bibfield  {author} {\bibinfo {author} {\bibfnamefont {R.~T.}\ \bibnamefont
  {Howie}}, \bibinfo {author} {\bibfnamefont {C.~L.}\ \bibnamefont
  {Guillaume}}, \bibinfo {author} {\bibfnamefont {T.}~\bibnamefont {Scheler}},
  \bibinfo {author} {\bibfnamefont {A.~F.}\ \bibnamefont {Goncharov}}, \ and\
  \bibinfo {author} {\bibfnamefont {E.}~\bibnamefont {Gregoryanz}},\ }\href
  {\doibase 10.1103/PhysRevLett.108.125501} {\bibfield  {journal} {\bibinfo
  {journal} {Phys. Rev. Lett.}\ }\textbf {\bibinfo {volume} {108}},\ \bibinfo
  {pages} {125501} (\bibinfo {year} {2012}{\natexlab{a}})}\BibitemShut
  {NoStop}%
\bibitem [{\citenamefont {Hemley}\ and\ \citenamefont
  {Mao}(1988)}]{PhysRevLett.61.857}%
  \BibitemOpen
  \bibfield  {author} {\bibinfo {author} {\bibfnamefont {R.~J.}\ \bibnamefont
  {Hemley}}\ and\ \bibinfo {author} {\bibfnamefont {H.~K.}\ \bibnamefont
  {Mao}},\ }\href {\doibase 10.1103/PhysRevLett.61.857} {\bibfield  {journal}
  {\bibinfo  {journal} {Phys. Rev. Lett.}\ }\textbf {\bibinfo {volume} {61}},\
  \bibinfo {pages} {857} (\bibinfo {year} {1988})}\BibitemShut {NoStop}%
\bibitem [{\citenamefont {Lorenzana}\ \emph {et~al.}(1989)\citenamefont
  {Lorenzana}, \citenamefont {Silvera},\ and\ \citenamefont
  {Goettel}}]{PhysRevLett.63.2080}%
  \BibitemOpen
  \bibfield  {author} {\bibinfo {author} {\bibfnamefont {H.~E.}\ \bibnamefont
  {Lorenzana}}, \bibinfo {author} {\bibfnamefont {I.~F.}\ \bibnamefont
  {Silvera}}, \ and\ \bibinfo {author} {\bibfnamefont {K.~A.}\ \bibnamefont
  {Goettel}},\ }\href {\doibase 10.1103/PhysRevLett.63.2080} {\bibfield
  {journal} {\bibinfo  {journal} {Phys. Rev. Lett.}\ }\textbf {\bibinfo
  {volume} {63}},\ \bibinfo {pages} {2080} (\bibinfo {year}
  {1989})}\BibitemShut {NoStop}%
\bibitem [{\citenamefont {Zha}\ \emph {et~al.}(2013)\citenamefont {Zha},
  \citenamefont {Liu}, \citenamefont {Ahart}, \citenamefont {Boehler},\ and\
  \citenamefont {Hemley}}]{PhysRevLett.110.217402}%
  \BibitemOpen
  \bibfield  {author} {\bibinfo {author} {\bibfnamefont {C.-S.}\ \bibnamefont
  {Zha}}, \bibinfo {author} {\bibfnamefont {Z.}~\bibnamefont {Liu}}, \bibinfo
  {author} {\bibfnamefont {M.}~\bibnamefont {Ahart}}, \bibinfo {author}
  {\bibfnamefont {R.}~\bibnamefont {Boehler}}, \ and\ \bibinfo {author}
  {\bibfnamefont {R.~J.}\ \bibnamefont {Hemley}},\ }\href {\doibase
  10.1103/PhysRevLett.110.217402} {\bibfield  {journal} {\bibinfo  {journal}
  {Phys. Rev. Lett.}\ }\textbf {\bibinfo {volume} {110}},\ \bibinfo {pages}
  {217402} (\bibinfo {year} {2013})}\BibitemShut {NoStop}%
\bibitem [{\citenamefont {Pickard}\ and\ \citenamefont
  {Needs}(2007)}]{Pickard07}%
  \BibitemOpen
  \bibfield  {author} {\bibinfo {author} {\bibfnamefont {C.~J.}\ \bibnamefont
  {Pickard}}\ and\ \bibinfo {author} {\bibfnamefont {R.~J.}\ \bibnamefont
  {Needs}},\ }\href@noop {} {\bibfield  {journal} {\bibinfo  {journal} {Nat.
  Phys.}\ }\textbf {\bibinfo {volume} {3}},\ \bibinfo {pages} {473} (\bibinfo
  {year} {2007})}\BibitemShut {NoStop}%
\bibitem [{\citenamefont {Azadi}\ and\ \citenamefont
  {K\"{u}hne}(2012)}]{Azadi12}%
  \BibitemOpen
  \bibfield  {author} {\bibinfo {author} {\bibfnamefont {S.}~\bibnamefont
  {Azadi}}\ and\ \bibinfo {author} {\bibfnamefont {T.~D.}\ \bibnamefont
  {K\"{u}hne}},\ }\href@noop {} {\bibfield  {journal} {\bibinfo  {journal}
  {JETP Lett.}\ }\textbf {\bibinfo {volume} {95}},\ \bibinfo {pages} {449}
  (\bibinfo {year} {2012})}\BibitemShut {NoStop}%
\bibitem [{\citenamefont {Azadi}\ \emph {et~al.}(2013)\citenamefont {Azadi},
  \citenamefont {Foulkes},\ and\ \citenamefont {K\"{u}hne}}]{Azadi13}%
  \BibitemOpen
  \bibfield  {author} {\bibinfo {author} {\bibfnamefont {S.}~\bibnamefont
  {Azadi}}, \bibinfo {author} {\bibfnamefont {W.~M.~C.}\ \bibnamefont
  {Foulkes}}, \ and\ \bibinfo {author} {\bibfnamefont {T.~D.}\ \bibnamefont
  {K\"{u}hne}},\ }\href@noop {} {\bibfield  {journal} {\bibinfo  {journal} {New
  J. Phys.}\ }\textbf {\bibinfo {volume} {15}},\ \bibinfo {pages} {113005}
  (\bibinfo {year} {2013})}\BibitemShut {NoStop}%
\bibitem [{\citenamefont {Azadi}\ and\ \citenamefont
  {Foulkes}(2013)}]{AzadiFoulkes13}%
  \BibitemOpen
  \bibfield  {author} {\bibinfo {author} {\bibfnamefont {S.}~\bibnamefont
  {Azadi}}\ and\ \bibinfo {author} {\bibfnamefont {W.~M.~C.}\ \bibnamefont
  {Foulkes}},\ }\href@noop {} {\bibfield  {journal} {\bibinfo  {journal} {Phys.
  Rev. B}\ }\textbf {\bibinfo {volume} {88}},\ \bibinfo {pages} {014115}
  (\bibinfo {year} {2013})}\BibitemShut {NoStop}%
\bibitem [{\citenamefont {Morales}\ \emph {et~al.}(2013)\citenamefont
  {Morales}, \citenamefont {McMahon}, \citenamefont {Pierleoni},\ and\
  \citenamefont {Ceperley}}]{Morales13b}%
  \BibitemOpen
  \bibfield  {author} {\bibinfo {author} {\bibfnamefont {M.~A.}\ \bibnamefont
  {Morales}}, \bibinfo {author} {\bibfnamefont {J.~M.}\ \bibnamefont
  {McMahon}}, \bibinfo {author} {\bibfnamefont {C.}~\bibnamefont {Pierleoni}},
  \ and\ \bibinfo {author} {\bibfnamefont {D.~M.}\ \bibnamefont {Ceperley}},\
  }\href {\doibase 10.1103/PhysRevB.87.184107} {\bibfield  {journal} {\bibinfo
  {journal} {Phys. Rev. B}\ }\textbf {\bibinfo {volume} {87}},\ \bibinfo
  {pages} {184107} (\bibinfo {year} {2013})}\BibitemShut {NoStop}%
\bibitem [{\citenamefont {Payne}\ \emph {et~al.}(1992)\citenamefont {Payne},
  \citenamefont {Teter}, \citenamefont {Allan}, \citenamefont {Arias},\ and\
  \citenamefont {Joannopoulos}}]{RevModPhys.64.1045}%
  \BibitemOpen
  \bibfield  {author} {\bibinfo {author} {\bibfnamefont {M.~C.}\ \bibnamefont
  {Payne}}, \bibinfo {author} {\bibfnamefont {M.~P.}\ \bibnamefont {Teter}},
  \bibinfo {author} {\bibfnamefont {D.~C.}\ \bibnamefont {Allan}}, \bibinfo
  {author} {\bibfnamefont {T.~A.}\ \bibnamefont {Arias}}, \ and\ \bibinfo
  {author} {\bibfnamefont {J.~D.}\ \bibnamefont {Joannopoulos}},\ }\href
  {\doibase 10.1103/RevModPhys.64.1045} {\bibfield  {journal} {\bibinfo
  {journal} {Rev. Mod. Phys.}\ }\textbf {\bibinfo {volume} {64}},\ \bibinfo
  {pages} {1045} (\bibinfo {year} {1992})}\BibitemShut {NoStop}%
\bibitem [{\citenamefont {K\"uhne}(2013)}]{CP2G}%
  \BibitemOpen
  \bibfield  {author} {\bibinfo {author} {\bibfnamefont {T.~D.}\ \bibnamefont
  {K\"uhne}},\ }\href {\doibase 10.1002/wcms.1176} {\bibfield  {journal}
  {\bibinfo  {journal} {WIREs Comput. Mol. Sci.}\ ,\ \bibinfo {pages} {1176}}
  (\bibinfo {year} {2013})}\BibitemShut {NoStop}%
\bibitem [{\citenamefont {Akahama}\ \emph
  {et~al.}(2010{\natexlab{a}})\citenamefont {Akahama}, \citenamefont
  {Kawamura}, \citenamefont {Hirao}, \citenamefont {Ohishi},\ and\
  \citenamefont {Takemura}}]{Akahama10}%
  \BibitemOpen
  \bibfield  {author} {\bibinfo {author} {\bibfnamefont {Y.}~\bibnamefont
  {Akahama}}, \bibinfo {author} {\bibfnamefont {H.}~\bibnamefont {Kawamura}},
  \bibinfo {author} {\bibfnamefont {N.}~\bibnamefont {Hirao}}, \bibinfo
  {author} {\bibfnamefont {Y.}~\bibnamefont {Ohishi}}, \ and\ \bibinfo {author}
  {\bibfnamefont {K.}~\bibnamefont {Takemura}},\ }\href
  {http://stacks.iop.org/1742-6596/215/i=1/a=012056} {\bibfield  {journal}
  {\bibinfo  {journal} {J. Phys.: Conf. Ser.}\ }\textbf {\bibinfo {volume}
  {215}},\ \bibinfo {pages} {012056} (\bibinfo {year}
  {2010}{\natexlab{a}})}\BibitemShut {NoStop}%
\bibitem [{\citenamefont {Zha}\ \emph {et~al.}(2012)\citenamefont {Zha},
  \citenamefont {Liu},\ and\ \citenamefont {Hemley}}]{Zha12}%
  \BibitemOpen
  \bibfield  {author} {\bibinfo {author} {\bibfnamefont {C.-S.}\ \bibnamefont
  {Zha}}, \bibinfo {author} {\bibfnamefont {Z.}~\bibnamefont {Liu}}, \ and\
  \bibinfo {author} {\bibfnamefont {R.~J.}\ \bibnamefont {Hemley}},\ }\href
  {\doibase 10.1103/PhysRevLett.108.146402} {\bibfield  {journal} {\bibinfo
  {journal} {Phys. Rev. Lett.}\ }\textbf {\bibinfo {volume} {108}},\ \bibinfo
  {pages} {146402} (\bibinfo {year} {2012})}\BibitemShut {NoStop}%
\bibitem [{\citenamefont {Bl\"ochl}(1994)}]{PhysRevB.50.17953}%
  \BibitemOpen
  \bibfield  {author} {\bibinfo {author} {\bibfnamefont {P.~E.}\ \bibnamefont
  {Bl\"ochl}},\ }\href {\doibase 10.1103/PhysRevB.50.17953} {\bibfield
  {journal} {\bibinfo  {journal} {Phys. Rev. B}\ }\textbf {\bibinfo {volume}
  {50}},\ \bibinfo {pages} {17953} (\bibinfo {year} {1994})}\BibitemShut
  {NoStop}%
\bibitem [{\citenamefont {Giannozzi}\ \emph {et~al.}(2009)\citenamefont
  {Giannozzi}, \citenamefont {Baroni}, \citenamefont {Bonini}, \citenamefont
  {Calandra}, \citenamefont {Car}, \citenamefont {Cavazzoni}, \citenamefont
  {Ceresoli}, \citenamefont {Chiarotti}, \citenamefont {Cococcioni},
  \citenamefont {Dabo}, \citenamefont {Corso}, \citenamefont {Fabris},
  \citenamefont {Fratesi}, \citenamefont {de~Gironcoli}, \citenamefont
  {Gebauer}, \citenamefont {Gerstmann}, \citenamefont {Gougoussis},
  \citenamefont {Kokalj}, \citenamefont {Lazzeri}, \citenamefont
  {Martin-Samos}, \citenamefont {Marzari}, \citenamefont {Mauri}, \citenamefont
  {Mazzarello}, \citenamefont {Paolini}, \citenamefont {Pasquarello},
  \citenamefont {Paulatto}, \citenamefont {Sbraccia}, \citenamefont {Scandolo},
  \citenamefont {Sclauzero}, \citenamefont {Seitsonen}, \citenamefont
  {Smogunov}, \citenamefont {Umari},\ and\ \citenamefont
  {Wentzcovitch}}]{Giannozzi09}%
  \BibitemOpen
  \bibfield  {author} {\bibinfo {author} {\bibfnamefont {P.}~\bibnamefont
  {Giannozzi}}, \bibinfo {author} {\bibfnamefont {S.}~\bibnamefont {Baroni}},
  \bibinfo {author} {\bibfnamefont {N.}~\bibnamefont {Bonini}}, \bibinfo
  {author} {\bibfnamefont {M.}~\bibnamefont {Calandra}}, \bibinfo {author}
  {\bibfnamefont {R.}~\bibnamefont {Car}}, \bibinfo {author} {\bibfnamefont
  {C.}~\bibnamefont {Cavazzoni}}, \bibinfo {author} {\bibfnamefont
  {D.}~\bibnamefont {Ceresoli}}, \bibinfo {author} {\bibfnamefont
  {G.}~\bibnamefont {Chiarotti}}, \bibinfo {author} {\bibfnamefont
  {M.}~\bibnamefont {Cococcioni}}, \bibinfo {author} {\bibfnamefont
  {I.}~\bibnamefont {Dabo}}, \bibinfo {author} {\bibfnamefont {A.~D.}\
  \bibnamefont {Corso}}, \bibinfo {author} {\bibfnamefont {S.}~\bibnamefont
  {Fabris}}, \bibinfo {author} {\bibfnamefont {G.}~\bibnamefont {Fratesi}},
  \bibinfo {author} {\bibfnamefont {S.}~\bibnamefont {de~Gironcoli}}, \bibinfo
  {author} {\bibfnamefont {R.}~\bibnamefont {Gebauer}}, \bibinfo {author}
  {\bibfnamefont {U.}~\bibnamefont {Gerstmann}}, \bibinfo {author}
  {\bibfnamefont {C.}~\bibnamefont {Gougoussis}}, \bibinfo {author}
  {\bibfnamefont {A.}~\bibnamefont {Kokalj}}, \bibinfo {author} {\bibfnamefont
  {M.}~\bibnamefont {Lazzeri}}, \bibinfo {author} {\bibfnamefont
  {L.}~\bibnamefont {Martin-Samos}}, \bibinfo {author} {\bibfnamefont
  {N.}~\bibnamefont {Marzari}}, \bibinfo {author} {\bibfnamefont
  {F.}~\bibnamefont {Mauri}}, \bibinfo {author} {\bibfnamefont
  {R.}~\bibnamefont {Mazzarello}}, \bibinfo {author} {\bibfnamefont
  {S.}~\bibnamefont {Paolini}}, \bibinfo {author} {\bibfnamefont
  {A.}~\bibnamefont {Pasquarello}}, \bibinfo {author} {\bibfnamefont
  {L.}~\bibnamefont {Paulatto}}, \bibinfo {author} {\bibfnamefont
  {C.}~\bibnamefont {Sbraccia}}, \bibinfo {author} {\bibfnamefont
  {S.}~\bibnamefont {Scandolo}}, \bibinfo {author} {\bibfnamefont
  {G.}~\bibnamefont {Sclauzero}}, \bibinfo {author} {\bibfnamefont {A.~P.}\
  \bibnamefont {Seitsonen}}, \bibinfo {author} {\bibfnamefont {A.}~\bibnamefont
  {Smogunov}}, \bibinfo {author} {\bibfnamefont {P.}~\bibnamefont {Umari}}, \
  and\ \bibinfo {author} {\bibfnamefont {R.~M.}\ \bibnamefont {Wentzcovitch}},\
  }\href@noop {} {\bibfield  {journal} {\bibinfo  {journal} {J. Phys.: Condens.
  Matter}\ }\textbf {\bibinfo {volume} {21}},\ \bibinfo {pages} {395502}
  (\bibinfo {year} {2009})}\BibitemShut {NoStop}%
\bibitem [{\citenamefont {Perdew}\ \emph {et~al.}(1996)\citenamefont {Perdew},
  \citenamefont {Burke},\ and\ \citenamefont {Ernzerhof}}]{Perdew96}%
  \BibitemOpen
  \bibfield  {author} {\bibinfo {author} {\bibfnamefont {J.~P.}\ \bibnamefont
  {Perdew}}, \bibinfo {author} {\bibfnamefont {K.}~\bibnamefont {Burke}}, \
  and\ \bibinfo {author} {\bibfnamefont {M.}~\bibnamefont {Ernzerhof}},\ }\href
  {\doibase 10.1103/PhysRevLett.77.3865} {\bibfield  {journal} {\bibinfo
  {journal} {Phys. Rev. Lett.}\ }\textbf {\bibinfo {volume} {77}},\ \bibinfo
  {pages} {3865} (\bibinfo {year} {1996})}\BibitemShut {NoStop}%
\bibitem [{\citenamefont {Monkhorst}\ and\ \citenamefont
  {Pack}(1976)}]{PhysRevB.13.5188}%
  \BibitemOpen
  \bibfield  {author} {\bibinfo {author} {\bibfnamefont {H.~J.}\ \bibnamefont
  {Monkhorst}}\ and\ \bibinfo {author} {\bibfnamefont {J.~D.}\ \bibnamefont
  {Pack}},\ }\href {\doibase 10.1103/PhysRevB.13.5188} {\bibfield  {journal}
  {\bibinfo  {journal} {Phys. Rev. B}\ }\textbf {\bibinfo {volume} {13}},\
  \bibinfo {pages} {5188} (\bibinfo {year} {1976})}\BibitemShut {NoStop}%
\bibitem [{\citenamefont {Hartwigsen}\ \emph {et~al.}(1998)\citenamefont
  {Hartwigsen}, \citenamefont {Goedecker},\ and\ \citenamefont
  {Hutter}}]{PhysRevB.58.3641}%
  \BibitemOpen
  \bibfield  {author} {\bibinfo {author} {\bibfnamefont {C.}~\bibnamefont
  {Hartwigsen}}, \bibinfo {author} {\bibfnamefont {S.}~\bibnamefont
  {Goedecker}}, \ and\ \bibinfo {author} {\bibfnamefont {J.}~\bibnamefont
  {Hutter}},\ }\href {\doibase 10.1103/PhysRevB.58.3641} {\bibfield  {journal}
  {\bibinfo  {journal} {Phys. Rev. B}\ }\textbf {\bibinfo {volume} {58}},\
  \bibinfo {pages} {3641} (\bibinfo {year} {1998})}\BibitemShut {NoStop}%
\bibitem [{\citenamefont {Perdew}\ and\ \citenamefont
  {Zunger}(1981)}]{PhysRevB.23.5048}%
  \BibitemOpen
  \bibfield  {author} {\bibinfo {author} {\bibfnamefont {J.~P.}\ \bibnamefont
  {Perdew}}\ and\ \bibinfo {author} {\bibfnamefont {A.}~\bibnamefont
  {Zunger}},\ }\href {\doibase 10.1103/PhysRevB.23.5048} {\bibfield  {journal}
  {\bibinfo  {journal} {Phys. Rev. B}\ }\textbf {\bibinfo {volume} {23}},\
  \bibinfo {pages} {5048} (\bibinfo {year} {1981})}\BibitemShut {NoStop}%
\bibitem [{\citenamefont {Car}\ and\ \citenamefont
  {Parrinello}(1985)}]{PhysRevLett.55.2471}%
  \BibitemOpen
  \bibfield  {author} {\bibinfo {author} {\bibfnamefont {R.}~\bibnamefont
  {Car}}\ and\ \bibinfo {author} {\bibfnamefont {M.}~\bibnamefont
  {Parrinello}},\ }\href {\doibase 10.1103/PhysRevLett.55.2471} {\bibfield
  {journal} {\bibinfo  {journal} {Phys. Rev. Lett.}\ }\textbf {\bibinfo
  {volume} {55}},\ \bibinfo {pages} {2471} (\bibinfo {year}
  {1985})}\BibitemShut {NoStop}%
\bibitem [{\citenamefont {K\"uhne}\ \emph {et~al.}(2007)\citenamefont
  {K\"uhne}, \citenamefont {Krack}, \citenamefont {Mohamed},\ and\
  \citenamefont {Parrinello}}]{PhysRevLett.98.066401}%
  \BibitemOpen
  \bibfield  {author} {\bibinfo {author} {\bibfnamefont {T.~D.}\ \bibnamefont
  {K\"uhne}}, \bibinfo {author} {\bibfnamefont {M.}~\bibnamefont {Krack}},
  \bibinfo {author} {\bibfnamefont {F.~R.}\ \bibnamefont {Mohamed}}, \ and\
  \bibinfo {author} {\bibfnamefont {M.}~\bibnamefont {Parrinello}},\ }\href
  {\doibase 10.1103/PhysRevLett.98.066401} {\bibfield  {journal} {\bibinfo
  {journal} {Phys. Rev. Lett.}\ }\textbf {\bibinfo {volume} {98}},\ \bibinfo
  {pages} {066401} (\bibinfo {year} {2007})}\BibitemShut {NoStop}%
\bibitem [{\citenamefont {Debernardi}\ and\ \citenamefont
  {Baroni}(1994)}]{Debernardi1994813}%
  \BibitemOpen
  \bibfield  {author} {\bibinfo {author} {\bibfnamefont {A.}~\bibnamefont
  {Debernardi}}\ and\ \bibinfo {author} {\bibfnamefont {S.}~\bibnamefont
  {Baroni}},\ }\href {\doibase http://dx.doi.org/10.1016/0038-1098(94)90654-8}
  {\bibfield  {journal} {\bibinfo  {journal} {Solid State Commun.}\ }\textbf
  {\bibinfo {volume} {91}},\ \bibinfo {pages} {813 } (\bibinfo {year}
  {1994})}\BibitemShut {NoStop}%
\bibitem [{\citenamefont {Baroni}\ \emph {et~al.}(2001)\citenamefont {Baroni},
  \citenamefont {de~Gironcoli}, \citenamefont {Dal~Corso},\ and\ \citenamefont
  {Giannozzi}}]{RevModPhys.73.515}%
  \BibitemOpen
  \bibfield  {author} {\bibinfo {author} {\bibfnamefont {S.}~\bibnamefont
  {Baroni}}, \bibinfo {author} {\bibfnamefont {S.}~\bibnamefont
  {de~Gironcoli}}, \bibinfo {author} {\bibfnamefont {A.}~\bibnamefont
  {Dal~Corso}}, \ and\ \bibinfo {author} {\bibfnamefont {P.}~\bibnamefont
  {Giannozzi}},\ }\href {\doibase 10.1103/RevModPhys.73.515} {\bibfield
  {journal} {\bibinfo  {journal} {Rev. Mod. Phys.}\ }\textbf {\bibinfo {volume}
  {73}},\ \bibinfo {pages} {515} (\bibinfo {year} {2001})}\BibitemShut
  {NoStop}%
\bibitem [{\citenamefont {Loubeyre}\ \emph {et~al.}(2002)\citenamefont
  {Loubeyre}, \citenamefont {Occelli},\ and\ \citenamefont
  {LeToullec}}]{Loubeyre02}%
  \BibitemOpen
  \bibfield  {author} {\bibinfo {author} {\bibfnamefont {P.}~\bibnamefont
  {Loubeyre}}, \bibinfo {author} {\bibfnamefont {F.}~\bibnamefont {Occelli}}, \
  and\ \bibinfo {author} {\bibfnamefont {R.}~\bibnamefont {LeToullec}},\
  }\href@noop {} {\bibfield  {journal} {\bibinfo  {journal} {Nature}\ }\textbf
  {\bibinfo {volume} {416}},\ \bibinfo {pages} {613} (\bibinfo {year}
  {2002})}\BibitemShut {NoStop}%
\bibitem [{\citenamefont {Akahama}\ \emph
  {et~al.}(2010{\natexlab{b}})\citenamefont {Akahama}, \citenamefont
  {Nishimura}, \citenamefont {Kawamura}, \citenamefont {Hirao}, \citenamefont
  {Ohishi},\ and\ \citenamefont {Takemura}}]{Akahama10a}%
  \BibitemOpen
  \bibfield  {author} {\bibinfo {author} {\bibfnamefont {Y.}~\bibnamefont
  {Akahama}}, \bibinfo {author} {\bibfnamefont {M.}~\bibnamefont {Nishimura}},
  \bibinfo {author} {\bibfnamefont {H.}~\bibnamefont {Kawamura}}, \bibinfo
  {author} {\bibfnamefont {N.}~\bibnamefont {Hirao}}, \bibinfo {author}
  {\bibfnamefont {Y.}~\bibnamefont {Ohishi}}, \ and\ \bibinfo {author}
  {\bibfnamefont {K.}~\bibnamefont {Takemura}},\ }\href {\doibase
  10.1103/PhysRevB.82.060101} {\bibfield  {journal} {\bibinfo  {journal} {Phys.
  Rev. B}\ }\textbf {\bibinfo {volume} {82}},\ \bibinfo {pages} {060101}
  (\bibinfo {year} {2010}{\natexlab{b}})}\BibitemShut {NoStop}%
\bibitem [{\citenamefont {Goncharov}\ \emph {et~al.}(2001)\citenamefont
  {Goncharov}, \citenamefont {Gregoryanz}, \citenamefont {Hemley},\ and\
  \citenamefont {Mao}}]{Goncharov01}%
  \BibitemOpen
  \bibfield  {author} {\bibinfo {author} {\bibfnamefont {A.~F.}\ \bibnamefont
  {Goncharov}}, \bibinfo {author} {\bibfnamefont {E.}~\bibnamefont
  {Gregoryanz}}, \bibinfo {author} {\bibfnamefont {R.~J.}\ \bibnamefont
  {Hemley}}, \ and\ \bibinfo {author} {\bibfnamefont {H.-K.}\ \bibnamefont
  {Mao}},\ }\href@noop {} {\bibfield  {journal} {\bibinfo  {journal} {Proc.
  Nat. Acad. Sci.}\ }\textbf {\bibinfo {volume} {98}},\ \bibinfo {pages}
  {14234} (\bibinfo {year} {2001})}\BibitemShut {NoStop}%
\bibitem [{\citenamefont {Hanfland}\ \emph {et~al.}(1993)\citenamefont
  {Hanfland}, \citenamefont {Hemley},\ and\ \citenamefont {Mao}}]{Hanfland93}%
  \BibitemOpen
  \bibfield  {author} {\bibinfo {author} {\bibfnamefont {M.}~\bibnamefont
  {Hanfland}}, \bibinfo {author} {\bibfnamefont {R.~J.}\ \bibnamefont
  {Hemley}}, \ and\ \bibinfo {author} {\bibfnamefont {H.-K.}\ \bibnamefont
  {Mao}},\ }\href {\doibase 10.1103/PhysRevLett.70.3760} {\bibfield  {journal}
  {\bibinfo  {journal} {Phys. Rev. Lett.}\ }\textbf {\bibinfo {volume} {70}},\
  \bibinfo {pages} {3760} (\bibinfo {year} {1993})}\BibitemShut {NoStop}%
\bibitem [{\citenamefont {Mazin}\ \emph {et~al.}(1997)\citenamefont {Mazin},
  \citenamefont {Hemley}, \citenamefont {Goncharov}, \citenamefont {Hanfland},\
  and\ \citenamefont {Mao}}]{Mazin97}%
  \BibitemOpen
  \bibfield  {author} {\bibinfo {author} {\bibfnamefont {I.~I.}\ \bibnamefont
  {Mazin}}, \bibinfo {author} {\bibfnamefont {R.~J.}\ \bibnamefont {Hemley}},
  \bibinfo {author} {\bibfnamefont {A.~F.}\ \bibnamefont {Goncharov}}, \bibinfo
  {author} {\bibfnamefont {M.}~\bibnamefont {Hanfland}}, \ and\ \bibinfo
  {author} {\bibfnamefont {H.-K.}\ \bibnamefont {Mao}},\ }\href {\doibase
  10.1103/PhysRevLett.78.1066} {\bibfield  {journal} {\bibinfo  {journal}
  {Phys. Rev. Lett.}\ }\textbf {\bibinfo {volume} {78}},\ \bibinfo {pages}
  {1066} (\bibinfo {year} {1997})}\BibitemShut {NoStop}%
\bibitem [{\citenamefont {Singh}\ \emph {et~al.}(2014)\citenamefont {Singh},
  \citenamefont {Azadi},\ and\ \citenamefont {K\"uhne}}]{SinghPhaseIV}%
  \BibitemOpen
  \bibfield  {author} {\bibinfo {author} {\bibfnamefont {R.}~\bibnamefont
  {Singh}}, \bibinfo {author} {\bibfnamefont {S.}~\bibnamefont {Azadi}}, \ and\
  \bibinfo {author} {\bibfnamefont {T.~D.}\ \bibnamefont {K\"uhne}},\
  }\href@noop {} {\bibfield  {journal} {\bibinfo  {journal} {to be published}\
  } (\bibinfo {year} {2014})}\BibitemShut {NoStop}%
\bibitem [{\citenamefont {Cui}\ \emph {et~al.}(1995)\citenamefont {Cui},
  \citenamefont {Chen},\ and\ \citenamefont {Silvera}}]{Cui95}%
  \BibitemOpen
  \bibfield  {author} {\bibinfo {author} {\bibfnamefont {L.}~\bibnamefont
  {Cui}}, \bibinfo {author} {\bibfnamefont {N.~H.}\ \bibnamefont {Chen}}, \
  and\ \bibinfo {author} {\bibfnamefont {I.~F.}\ \bibnamefont {Silvera}},\
  }\href {\doibase 10.1103/PhysRevB.51.14987} {\bibfield  {journal} {\bibinfo
  {journal} {Phys. Rev. B}\ }\textbf {\bibinfo {volume} {51}},\ \bibinfo
  {pages} {14987} (\bibinfo {year} {1995})}\BibitemShut {NoStop}%
\bibitem [{\citenamefont {Howie}\ \emph
  {et~al.}(2012{\natexlab{b}})\citenamefont {Howie}, \citenamefont {Scheler},
  \citenamefont {Guillaume},\ and\ \citenamefont {Gregoryanz}}]{Howie12a}%
  \BibitemOpen
  \bibfield  {author} {\bibinfo {author} {\bibfnamefont {R.~T.}\ \bibnamefont
  {Howie}}, \bibinfo {author} {\bibfnamefont {T.}~\bibnamefont {Scheler}},
  \bibinfo {author} {\bibfnamefont {C.~L.}\ \bibnamefont {Guillaume}}, \ and\
  \bibinfo {author} {\bibfnamefont {E.}~\bibnamefont {Gregoryanz}},\ }\href
  {\doibase 10.1103/PhysRevB.86.214104} {\bibfield  {journal} {\bibinfo
  {journal} {Phys. Rev. B}\ }\textbf {\bibinfo {volume} {86}},\ \bibinfo
  {pages} {214104} (\bibinfo {year} {2012}{\natexlab{b}})}\BibitemShut
  {NoStop}%
\bibitem [{\citenamefont {Liu}\ and\ \citenamefont {Ma}(2013)}]{Liu13}%
  \BibitemOpen
  \bibfield  {author} {\bibinfo {author} {\bibfnamefont {H.}~\bibnamefont
  {Liu}}\ and\ \bibinfo {author} {\bibfnamefont {Y.}~\bibnamefont {Ma}},\
  }\href {\doibase 10.1103/PhysRevLett.110.025903} {\bibfield  {journal}
  {\bibinfo  {journal} {Phys. Rev. Lett.}\ }\textbf {\bibinfo {volume} {110}},\
  \bibinfo {pages} {025903} (\bibinfo {year} {2013})}\BibitemShut {NoStop}%
\bibitem [{\citenamefont {Magd\ifmmode~\u{a}\else \u{a}\fi{}u}\ and\
  \citenamefont {Ackland}(2013)}]{Magdau13}%
  \BibitemOpen
  \bibfield  {author} {\bibinfo {author} {\bibfnamefont {I.~B.}\ \bibnamefont
  {Magd\ifmmode~\u{a}\else \u{a}\fi{}u}}\ and\ \bibinfo {author} {\bibfnamefont
  {G.~J.}\ \bibnamefont {Ackland}},\ }\href {\doibase
  10.1103/PhysRevB.87.174110} {\bibfield  {journal} {\bibinfo  {journal} {Phys.
  Rev. B}\ }\textbf {\bibinfo {volume} {87}},\ \bibinfo {pages} {174110}
  (\bibinfo {year} {2013})}\BibitemShut {NoStop}%
\bibitem [{\citenamefont {Kaxiras}\ \emph {et~al.}(1991)\citenamefont
  {Kaxiras}, \citenamefont {Broughton},\ and\ \citenamefont
  {Hemley}}]{PhysRevLett.67.1138}%
  \BibitemOpen
  \bibfield  {author} {\bibinfo {author} {\bibfnamefont {E.}~\bibnamefont
  {Kaxiras}}, \bibinfo {author} {\bibfnamefont {J.}~\bibnamefont {Broughton}},
  \ and\ \bibinfo {author} {\bibfnamefont {R.~J.}\ \bibnamefont {Hemley}},\
  }\href {\doibase 10.1103/PhysRevLett.67.1138} {\bibfield  {journal} {\bibinfo
   {journal} {Phys. Rev. Lett.}\ }\textbf {\bibinfo {volume} {67}},\ \bibinfo
  {pages} {1138} (\bibinfo {year} {1991})}\BibitemShut {NoStop}%
\bibitem [{\citenamefont {Perdew}\ and\ \citenamefont
  {Levy}(1983)}]{PhysRevLett.51.1884}%
  \BibitemOpen
  \bibfield  {author} {\bibinfo {author} {\bibfnamefont {J.~P.}\ \bibnamefont
  {Perdew}}\ and\ \bibinfo {author} {\bibfnamefont {M.}~\bibnamefont {Levy}},\
  }\href {\doibase 10.1103/PhysRevLett.51.1884} {\bibfield  {journal} {\bibinfo
   {journal} {Phys. Rev. Lett.}\ }\textbf {\bibinfo {volume} {51}},\ \bibinfo
  {pages} {1884} (\bibinfo {year} {1983})}\BibitemShut {NoStop}%
\bibitem [{\citenamefont {St\"adele}\ and\ \citenamefont
  {Martin}(2000)}]{Stadele00}%
  \BibitemOpen
  \bibfield  {author} {\bibinfo {author} {\bibfnamefont {M.}~\bibnamefont
  {St\"adele}}\ and\ \bibinfo {author} {\bibfnamefont {R.~M.}\ \bibnamefont
  {Martin}},\ }\href {\doibase 10.1103/PhysRevLett.84.6070} {\bibfield
  {journal} {\bibinfo  {journal} {Phys. Rev. Lett.}\ }\textbf {\bibinfo
  {volume} {84}},\ \bibinfo {pages} {6070} (\bibinfo {year}
  {2000})}\BibitemShut {NoStop}%
\bibitem [{\citenamefont {{Barbee III}}\ \emph {et~al.}(1989)\citenamefont
  {{Barbee III}}, \citenamefont {Garcia},\ and\ \citenamefont
  {Cohen}}]{HydrogenHTC}%
  \BibitemOpen
  \bibfield  {author} {\bibinfo {author} {\bibfnamefont {T.~W.}\ \bibnamefont
  {{Barbee III}}}, \bibinfo {author} {\bibfnamefont {A.}~\bibnamefont
  {Garcia}}, \ and\ \bibinfo {author} {\bibfnamefont {M.~L.}\ \bibnamefont
  {Cohen}},\ }\href@noop {} {\bibfield  {journal} {\bibinfo  {journal}
  {Nature}\ }\textbf {\bibinfo {volume} {340}},\ \bibinfo {pages} {369}
  (\bibinfo {year} {1989})}\BibitemShut {NoStop}%
\bibitem [{\citenamefont {Richardson}\ and\ \citenamefont
  {Ashcroft}(1997)}]{PhysRevLett.78.118}%
  \BibitemOpen
  \bibfield  {author} {\bibinfo {author} {\bibfnamefont {C.~F.}\ \bibnamefont
  {Richardson}}\ and\ \bibinfo {author} {\bibfnamefont {N.~W.}\ \bibnamefont
  {Ashcroft}},\ }\href {\doibase 10.1103/PhysRevLett.78.118} {\bibfield
  {journal} {\bibinfo  {journal} {Phys. Rev. Lett.}\ }\textbf {\bibinfo
  {volume} {78}},\ \bibinfo {pages} {118} (\bibinfo {year} {1997})}\BibitemShut
  {NoStop}%
\end{thebibliography}
%

\end{document}